\documentclass[aps,prb,twocolumn]{revtex4}  
\usepackage{graphicx}
\usepackage{dcolumn}
\usepackage{bm}
\usepackage{amsmath}
\newcommand{\comment}[1]{}

\begin{document}
\renewcommand{\theequation}{\arabic{section}.\arabic{equation}}


\title{Thermodynamic Stability of Nanobubbles}

\author{Phil Attard}

\affiliation{\emph{phil.attard1@gmail.com}, Sydney, Australia}

\date{3.14\,15}

\begin{abstract}
The observed stability of nanobubbles contradicts
the well-known result in classical nucleation theory,
that the critical radius is both microscopic and thermodynamically unstable.
Here nanoscopic stability is shown
to be the combined result of  two non-classical mechanisms.
It is shown that the surface tension decreases with increasing supersaturation,
and that this gives a nanoscopic critical radius.
Whilst neither a free spherical bubble
nor a  hemispherical bubble mobile on an hydrophobic surface are stable,
it is shown that an immobilized hemispherical bubble
with a pinned contact rim is stable
and that the total entropy is a maximum at the critical radius.
\end{abstract}

\maketitle

%
\section{Introduction: Nanobubble Trouble}
\setcounter{equation}{0} 
%


Experimental evidence for nanobubbles was first published in 1994,
\cite{Parker94}
and since then their existence has been confirmed
from various features of the measured forces
between hydrophobic surfaces,
\cite{Carambassis98,Tyrrell01,Tyrrell02}
including a reduced attraction in de-aerated water,
\cite{Wood95,Meagher94,Considine99,Mahnke99,Ishida00,Meyer05,Stevens05}
and from images obtained
with tapping mode atomic force microscopy.
\cite{Tyrrell01,Tyrrell02,Ishida00,Holmberg03,Simonsen04,%
Zhang04,Zhang06a,Zhang06b,Yang07}
For a recent review of theory and experiment,
see Ref.~\onlinecite{Attard14}.

In the fields of nucleation science and bubble research,
three concepts are important.\cite{Rowlinson82,Oxtoby98}
First, the critical radius,
which is the point at which the derivatives of the total entropy
(equivalently the free energy) vanish.
Second, the Laplace-Young equation,
which says that the internal gas pressure of the bubble
is larger than the pressure of the surrounding liquid
by an amount equal to twice the surface tension divided by radius.
Third, for the case of heterogenous nucleation and bubbles adhered to surfaces,
the Young equation,
which says that the cosine of the contact angle
is equal to the difference in surface energies of the solid divided
by the liquid-vapor surface tension.

The trouble with nanobubbles is that
\begin{enumerate}
\item
the critical radius is an unstable equilibrium point
(it is a saddle point),
whereas nanobubbles appear to be stable
\item
the conventional critical radius for realistic parameters
is an order of magnitude or so larger
than the typical nanobubble radius of curvature
\item
the Laplace-Young equation predicts that typical nanobubbles
have an internal gas pressure
one or two orders of magnitude larger than atmospheric pressure
\item
the measured contact angle for  typical nanobubbles
is significantly larger than the contact angle
measured for macroscopic droplets on the same surface
\end{enumerate}
Any explanation for the existence of nanobubbles must reconcile
these four points with conventional thermodynamic principles
and realistic physical models.

The first point is such a strong qualitative contradiction
with the experimentally observed stability (hours or days) of nanobubbles
that it calls into question conventional nucleation thermodynamics itself.
The second point represents a quantitative discrepancy
between the length scales for bubbles predicted by
thermodynamics using conventional parameter values
and the measured length scales of nanobubbles.
The third point
says that there is a strong thermodynamic gradient
between the gas inside the nanobubble and the atmosphere,
which should lead to the runaway dissolution
of a nanobubble on time-scales orders of magnitude
shorter than the experimentally observed lifetimes.
The fourth point again appears to prove
that conventional macroscopic thermodynamics,
both bulk and surface, is inapplicable on nanoscopic length scales,
which broad point
is at variance with diverse experimental measurements in other systems,
molecular-level computer simulation data,
and  conventional wisdom.
These four points are unlikely to be independent of each other.

\begin{figure}[t!]
\centerline{
\resizebox{8.5cm}{!}{ \includegraphics*{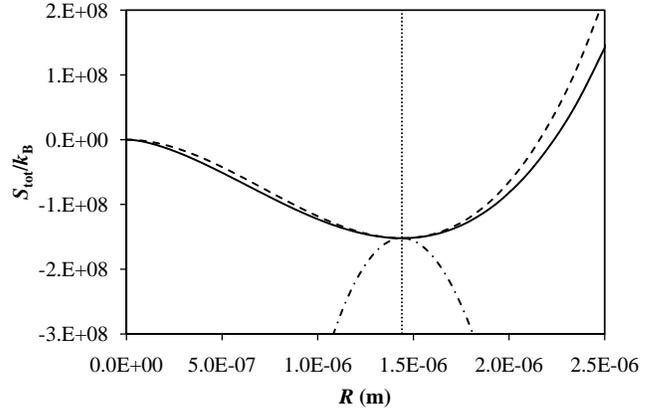} } }
\caption{\label{Fig:b2}
Total entropy of a spherical air bubble in water
as a function of radius for a supersaturation ratio of 2
and a surface tension of $0.072\,$N/m.
The solid curve follows the mechanical equilibrium  path,
$\partial S_\mathrm{tot}/\partial R = 0$,
the dashed curve follows the diffusive equilibrium  path,
$\partial S_\mathrm{tot}/\partial N = 0$,
and the dash-dotted curve follows the constant number path,
$N = N_\mathrm{crit}$.
The dotted line marks the critical radius.
}
\end{figure}

Figure~\ref{Fig:b2} shows the total entropy
for a nucleating spherical bubble.
A stable equilibrium point corresponds to a local maximum of the entropy.
(In this article all thermodynamic results are cast in terms
of entropy. These can be translated into free energy
by multiplying throughout by minus the temperature.
Hence maximum total entropy is the same as minimum free energy.)
It can be seen that when the growth of the bubble occurs
along a path of either mechanical or diffusive equilibrium,
the total entropy  is a minimum  at the critical radius.
However if the change in size occurs at either constant number (shown)
or constant radius (not shown),
then the total entropy is a maximum at the critical radius.
Hence the total entropy
has a saddle point at the critical radius,
which means that it is a point of unstable equilibrium.
The critical radius for these parameters
is about an order of magnitude larger than the radius of curvature
typically measured for nanobubbles.

\label{para:unstable}
The physical reason for the instability of the critical radius
is readily understood.
At the critical radius the diffusive (number)
and mechanical (radius) gradients are zero.
A fluctuation to a larger radius along the path of mechanical equilibrium
(mechanical equilibrium occurs faster than diffusive equilibrium
with the reservoir beyond the immediate vicinity of the surface)
reduces the internal pressure and hence the gas chemical potential
inside the bubble. There is now a thermodynamic gradient
that forces dissolved gas from the reservoir into the bubble,
which increases the internal pressure.
This creates a mechanical force imbalance,
which acts to increase the volume of the bubble
at the expense of the volume of the reservoir.
This  increases the radius,
magnifying the original fluctuation.
The opposite occurs if the initial fluctuation
decreases the radius,
in which case the bubble is driven to zero radius.

An apparently trivial but essential point
is easily overlooked in this explanation:
The instability at the critical radius
relies upon the fact that the volume of the bubble
increases with increasing radius.
It will turn out that this simple geometric fact
holds the key to understanding
the thermodynamic stability of nanobubbles.

The purpose of this paper is to show
what it takes to make the critical radius
of a bubble  both stable and nanoscopic.

\section{Conventional Nucleation Thermodynamics}
\setcounter{equation}{0} \setcounter{subsubsection}{0}
%

Consider an air bubble in water.
Only the air need be explicitly accounted for,
as the water solvent can be considered always at equilibrium
(see Appendix~\ref{Sec:2cpt}).
The total entropy is taken to be the sum of three terms:
the bulk entropy of the sub-system, which is the bubble,
the bulk entropy of the  reservoir, which is the air dissolved in water,
and the interfacial entropy of their boundary,\cite{Attard02}
\begin{eqnarray}
\lefteqn{
S_\mathrm{tot}(N,V,T)
} \nonumber \\
& = &
S_\mathrm{s}(N,V,T)
+ S_\mathrm{r}(N_\mathrm{r},V_\mathrm{r},T)
- \frac{\gamma A}{T }
\nonumber \\  & = &
S_\mathrm{s}(N,V,T)
- \frac{ p V}{T}
+ \frac{ \mu N}{T}
- \frac{4 \pi \gamma R^2}{T } .
\end{eqnarray}
The total system is in thermal equilibrium at temperature $T$.
Here $N$ is the number of molecules in the bubble,
$V=4 \pi R^3/3$ is its volume,
$A=4 \pi R^2$ is its area,
and $R$ is its radius.
The pressure of the reservoir is $p$,
which  equals atmospheric pressure,
the chemical potential  of the reservoir is $\mu$,
and the liquid-vapor surface tension is $\gamma$.
As shown in Appendix~\ref{Sec:2cpt},
it would be more precise to set the external reservoir pressure
equal to the partial vapor pressure of air at saturation,
$p = p^\dag \approx 0.8 p_\mathrm{atm}$,
but for simplicity this is not done here.

In the capillarity approximation,
the bubble is taken to be an ideal gas\cite{Rowlinson82,Attard02}
\begin{equation}
S_\mathrm{s}(N,V,T)
=
k_\mathrm{B} N \left[ 1 - \ln \frac{N \Lambda^3}{V} \right] ,
\end{equation}
where $k_\mathrm{B} $ is Boltzmann's constant
and $\Lambda$ is the thermal wave length.
Although the entropy of a real gas  differs from this quantitatively,
the ideal gas model is not expected to affect the results qualitatively.

A bubble can only be in equilibrium if the water is supersaturated with air
(see below).
The supersaturation ratio, $s$, is the ratio of the actual concentration
of dissolved air to the saturation  concentration,
which equals the ratio of the air pressures that would give
the two concentrations, $s= p_\mathrm{ss}/p^\dag$,
where the dagger here and below denotes saturation.
The chemical potential can be obtained from the supersaturation ratio
by the standard ideal gas result\cite{Attard02}
\begin{equation}
\mu
= k_\mathrm{B}T \ln \frac{s p^\dag\Lambda^3}{k_\mathrm{B}T} .
\end{equation}
In this work the saturation air pressure will be taken
to be atmospheric pressure, $p^\dag = p$.
This neglects the vapor pressure of water,
which is about one fifth of an atmosphere.

The number derivative is
\begin{equation}
\frac{\partial S_\mathrm{tot}(N,V,T)}{\partial N}
=
- \frac{\mu_\mathrm{s} }{T}
+ \frac{\mu }{T} ,
\end{equation}
which vanishes when $\overline \mu_\mathrm{s} = \mu$.
Using the ideal gas equation this is
\begin{equation}
\overline \rho
=
 \Lambda^{-3} e^{\mu/k_\mathrm{B}T}
=
\frac{s p}{k_\mathrm{B}T} ,
\end{equation}
where $\rho = N/V$ is the number density of air
inside the bubble.
This is the equation for diffusive equilibrium
between the air inside the bubble and the air dissolved in the water.

The volume derivative is
\begin{equation}
\frac{\partial S_\mathrm{tot}(N,V,T)}{\partial V} =
\frac{p_\mathrm{s} }{T}
- \frac{p }{T}
- \frac{2 \gamma }{TR} .
\end{equation}
This vanishes when
\begin{equation}
\overline p_\mathrm{s} =
p + \frac{2 \gamma }{ R}  ,
\end{equation}
which is the  Laplace-Young equation.
This is the equation for mechanical equilibrium
for the bubble.

When the bubble is in  mechanical equilibrium,
the  Laplace-Young equation says that
the internal pressure exceeds the external pressure.
Since the pressure is a monotonic function of the chemical potential,
and since the external pressure is equal to the saturation vapor pressure,
the chemical potential of the air inside the bubble
must exceed the saturation chemical potential.
This is the reason that a bubble in mechanical equilibrium
can be in diffusive equilibrium only with a supersaturated solution.
\cite{Moody02}

The critical radius is the point
at which diffusive and mechanical equilibrium simultaneously hold.
Using the the ideal gas equation of state,
$p_\mathrm{s} = \rho k_\mathrm{B}T$,
and substituting the diffusive equilibrium equation
into the mechanical  equilibrium equation,
the critical radius is
\begin{equation}
R_\mathrm{crit}
=
\frac{2 \gamma }{ (s-1) p } .
\end{equation}
The critical radius scales linearly with the surface tension
and decreases with increasing supersaturation ratio.
The depth of the entropy minimum
(equivalently, height of the free energy barrier)
scales with the cube of the critical radius,
$S_\mathrm{tot,crit}= - k_\mathrm{B} (s-1) \rho^\dag V_\mathrm{crit}/2$.

The numerical results in Fig.~\ref{Fig:b2}
show that the critical radius is a point of unstable equilibrium,
which is of course well-known in classical nucleation theory.
\cite{Rowlinson82,Oxtoby98}
See Appendix~\ref{Sec:Unstab-Rc} for a mathematical proof of the result.
Also this, the conventional formula for the critical radius, would require
a supersaturation ratio $s = $10--100
(equivalently, water in equilibrium with air at 10--100 atmospheres)
in order to produce a critical radius $R_\mathrm{crit}=10^{-7}$--$10^{-8}\,$m,
which are the radii measured for nanobubbles.

%
\section{State Dependent Surface Tension}
\setcounter{equation}{0} \setcounter{subsubsection}{0}
%

As explained above,
if a bubble is in diffusive equilibrium with the water phase,
then the solution must be supersaturated with air.\cite{Moody02}
The effects of supersaturation on the surface tension
are neglected in the classical theory,
but in fact these effects can be quite dramatic.

The barrier to bubble growth
is the depth of the entropy at the critical radius,
which scales with the cube of the surface tension.
At the spinodal supersaturation,
there is no barrier to bubble growth,
which means that the surface tension must vanish.\cite{Oxtoby98}
Hence the surface tension must decrease with increasing supersaturation.
\cite{Moody03,Moody04,He05}

This general thermodynamic result is consistent
with what is known experimentally  from high pressure measurements,
namely that the surface tension of water is lower
in the presence of oxygen or nitrogen and water vapor
than in the presence of pure water vapor.\cite{Massoudi74}

Computer simulations of the supersaturated liquid-vapor interface
show that to a reasonable approximation the decrease is linear,
\cite{Moody03,Moody04,He05}
\begin{eqnarray}
\gamma(s)
& = &
\gamma^\dag \frac{s^\ddag- s}{s^\ddag-1}
\nonumber \\ & = &
\gamma^\dag
\frac{p_\mathrm{g}^\ddag - p_\mathrm{g}}{p_\mathrm{g}^\ddag-p_\mathrm{g}^\dag}.
\end{eqnarray}
Here $\gamma^\dag = 0.072\,$N/m
is the saturation or coexistence surface tension
(it was denoted plain $\gamma$ above),
$ s \equiv p_\mathrm{g}/p_\mathrm{g}^\dag$
is the reservoir supersaturation ratio,
which is the ratio of the supersaturation air  vapor pressure
to the saturation air pressure,
and the dagger and double dagger denote
the coexistence and the spinodal values, respectively.
Obviously, $s^\ddag > s^\dag = 1$ and
$p_\mathrm{g}^\ddag>p_\mathrm{g}^\dag$.
Simulations of a Lennard-Jones fluid at various temperatures
found that the surface tension was well-fitted by this
with  spinodal supersaturation ratios in the range $s^\ddag =$ 3--6.
\cite{Moody03,Moody04,He05}

This surface tension may be called \emph{the} supersaturated surface tension.
It has a constant value independent of the radius of the bubble
and dependent only on the supersaturation ratio of the reservoir.
(the concentration of air dissolved in the water
divided by the saturation concentration).

If one uses this reservoir state-dependent  surface tension,
then all of the results of classical nucleation theory hold,
with the value of the surface tension being replaced
by the supersaturated value,
$\gamma^\dag \Rightarrow \gamma(s)$.
This means that both the critical radius,
which scales linearly with the surface tension,
and the nucleation barrier,
which scales cubicly with the surface tension,
are reduced from their classical values.

\begin{figure}[t!]
\centerline{
\resizebox{8.5cm}{!}{ \includegraphics*{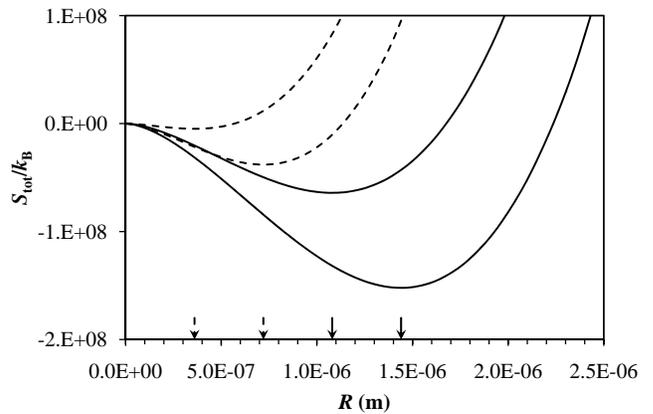} } }
\caption{\label{Fig:bg2}
Total entropy of a spherical air bubble in water
as a function of radius following the mechanical equilibrium  path.
The solid curves are for a supersaturation ratio of $s=2$
and the dashed curves are for $s=3$.
The lower curves of each pair
use the saturation surface tension $\gamma^\dag = 0.072\,$N/m,
and the upper curves  of each pair use the supersaturated surface tension
(spinodal supersaturation ratio $ s^\ddag = 5$),
$\gamma(2) = 0.054\,$N/m,
and $\gamma(3)=0.036\,$N/m.
The arrows mark the respective critical radii (local entropy minimum).
}
\end{figure}

\begin{figure}[t!]
\centerline{
\resizebox{8.5cm}{!}{ \includegraphics*{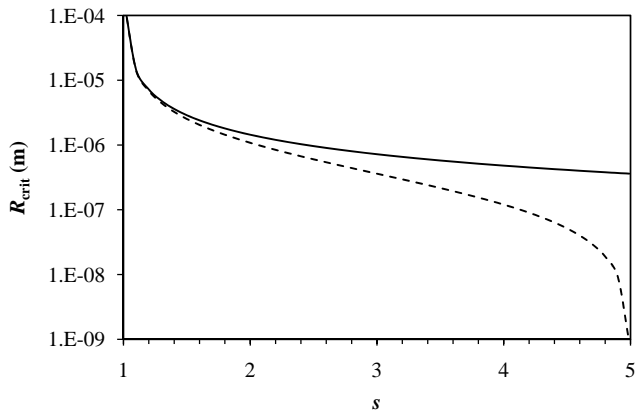} } }
\caption{\label{Fig:bg3}
Critical radius (log scale) as a function of supersaturation
for an air bubble in water.
The solid curve uses the saturation  surface tension,
$\gamma^\dag = 0.072\,$N/m,
and the dashed curve uses the supersaturation surface tension,
$\gamma(s)$, with $s^\ddag = 5$.
}
\end{figure}

Figure~\ref{Fig:bg2}
shows the effect of the supersaturated surface tension
on the total entropy of a bubble.
It can be seen that the critical radius and the nucleation barrier
are indeed reduced from their conventional values.
The critical radius is nanoscopic for realistic parameter values,
and would move to even smaller values
for larger reservoir supersaturation ratios,
or if the spinodal supersaturation ratio had a smaller value.
The critical radius is plotted as a function of supersaturation
in Fig.~\ref{Fig:bg3}.

Explicit comparison can be made with measurements on nanobubbles.
A supersaturation ratio of
$s=3.6$ and a spinodal supersaturation of $s^\ddag = 5$
gives a supersaturation surface tension of
$\gamma(s) = 0.026\,$N/m
and a critical radius of $R_\mathrm{crit}=200\,$nm.
Alternatively,
$s=4.4$ gives $\gamma(s) = 0.010\,$N/m
and a critical radius of $R_\mathrm{crit}=60\,$nm.
These are comparable to values obtained
from measurements on nanobubbles,
specifically from analysis of the difference
between the nanoscopic and macroscopic contact angles,\cite{Attard14}
with the former being obtained from
from the tapping mode image profiles of nanobubbles.\cite{Tyrrell01,Tyrrell02}
In one series of experiments
the average radius of curvature was $R=200\,$nm,
and a surface tension of $\gamma(s) = 0.015\,$N/m was deduced
from the contact angle.
From another series,  $R=60\,$nm and $\gamma(s) = 0.019\,$N/m
were obtained.
An alternative interpretation of these results
is discussed in the conclusion to the paper.

From the curves in Fig.~\ref{Fig:bg2}
it can be seen that the critical radius remains an entropy minimum
when the supersaturated surface tension is used.
Hence the nanobubble remains unstable.
In Appendix~\ref{Sec:Local-Super-Sat}
the effect of using the local supersaturation
(ie.\ the supersaturation in the solution adjacent
to the air-water interface,
which is in diffusive equilibrium with the air inside the bubble),
which is greater than the reservoir supersaturation, is discussed.
Even in this case,
and for a general non-linear model of the supersaturated surface tension,
it is proven that the nanobubble is always unstable.
Hence in general the supersaturated surface tension alone
cannot give a stable nanobubble.

Despite the lack of stability,
the decrease in surface tension
with increasing supersaturation appears to solve
one of the two major problems
that nanobubbles pose for conventional thermodynamics.
As has been previously argued,\cite{Moody03}
it explains how the equilibrium radius comes to be of nanoscopic dimensions.

%
\section{Hemispherical Bubble on a Surface}
\setcounter{equation}{0} \setcounter{subsubsection}{0}
%

\subsection{Mobile Hemispherical Bubble}

For a bubble on a solid surface,
the optimum exterior contact angle is given by the Young equation
\begin{equation}
\gamma_\mathrm{sl} =
\gamma_\mathrm{sg} + \gamma_\mathrm{lg} \cos (\pi-\overline \theta_\mathrm{c})
, \mbox{ or }
\cos \overline \theta_\mathrm{c}
= \frac{\gamma_\mathrm{sg}-\gamma_\mathrm{sl}}{\gamma_\mathrm{lg}}.
\end{equation}
The liquid-vapor surface tension $\gamma_\mathrm{lg}$
was variously denoted $\gamma$, or $\gamma^\dag$, or $\gamma(s)$ above.
For a hydrophobic surface, which is the present concern,
the  surface energies of the solid interfaces satisfy
$\gamma_\mathrm{sg}<\gamma_\mathrm{sl}$,
or $\overline \theta_\mathrm{c} > \pi/2$.
Define $\Delta \gamma \equiv \gamma_\mathrm{sg}-\gamma_\mathrm{sl} < 0$.

It should be noted that
the difference in the solid surface energies is assumed fixed
with a value determined from measurements on macroscopic water droplets
on the given surface and the saturation surface tension.
Consequently the optimum contact angle for nanobubbles,
which is given by the Young equation,
increases with increasing supersaturation ratio
since the liquid-vapor surface tension
decreases with  increasing supersaturation ratio.
This notion is the origin of the values of the surface tension
given above,
which were deduced from the measured contact angle for nanobubbles.
\cite{Attard14}

Consider an air bubble with radius of curvature $R$,
apex height above the surface of $h$,
and contact rim radius $r$.
The contact rim radius and apex height are related by
\begin{equation}
r^2 + h^2 - 2 R h = 0.
\end{equation}
Hence $r =\sqrt{2Rh-h^2}$ or $h = R- \sqrt{R^2-r^2}$.

The interior contact angle is $\theta' = \pi - \theta$,
and the angle between the surface and the curvature radius
at the contact rim is
$\phi = (\pi/2)-\theta' = \theta - \pi/2$.
One has  $(R-h)/R = \sin \phi = -\cos \theta$,
or $\cos \theta = -1 + h/R$.

The volume of the bubble is
\begin{equation}
V =
\pi \left[ R h^2 - \frac{1}{3} h^3  \right] ,
\end{equation}
and the area of the liquid-vapor interface is
\begin{equation}
A_\mathrm{lg} =
2 \pi R \left[ R  - \sqrt{ R^2 - r^2}  \right] = 2 \pi R h.
\end{equation}
The area of the solid-gas interface is of course
$A_\mathrm{sg} = \pi r^2 = \pi (2 R h - h^2) $.

The total entropy is
\begin{eqnarray}
S_\mathrm{total}
& = &
S_\mathrm{s}(N,V,T)
+ \frac{\mu}{T} N - \frac{p}{T}V
\nonumber \\ & & \mbox{ }
- \frac{\gamma_\mathrm{lg}  }{T}A_\mathrm{lg}
- \frac{\Delta \gamma}{T} A_\mathrm{sg}  .
\end{eqnarray}
Again one can use the ideal gas entropy for the bubble,
$S_\mathrm{s}(N,V,T) = Nk_\mathrm{B} [ 1 - \ln N \Lambda^3/V ]$.
The total entropy is most conveniently written
as a function of $N$, $R$, and either $h$ or $r$.

Using $S_\mathrm{total}(N,R,h|T)$,
the derivatives are
\begin{equation} \label{Eq:ds/dN}
\frac{\partial S_\mathrm{total}}{\partial N}
= \frac{-\mu_\mathrm{s}}{T} + \frac{\mu}{T} ,
\end{equation}
\begin{eqnarray}\label{Eq:ds/dR}
\frac{\partial S_\mathrm{total}}{\partial R}
& = &
\left[ \frac{p_\mathrm{s}}{T} - \frac{p}{T} \right]
\frac{\partial V(R,h)}{\partial R}
\nonumber \\  & & \mbox{ }
- \frac{\gamma_\mathrm{lg}  }{T}
\frac{\partial A_\mathrm{lg}(R,h)}{\partial R}
- \frac{\Delta \gamma}{T}
\frac{\partial A_\mathrm{sg}(R,h)}{\partial R}
\nonumber \\  & = &
\left[ \frac{p_\mathrm{s}}{T} - \frac{p}{T} \right]
\pi h^2
- \frac{\gamma_\mathrm{lg}  + \Delta \gamma }{T}
2 \pi h ,
\end{eqnarray}
and
\begin{eqnarray}\label{Eq:ds/dh}
\frac{\partial S_\mathrm{total}}{\partial h}
& = &
\left[ \frac{p_\mathrm{s}}{T} - \frac{p}{T} \right]
\frac{\partial V(R,h)}{\partial h}
\nonumber \\ & & \mbox{ }
- \frac{\gamma_\mathrm{lg}  }{T}
\frac{\partial A_\mathrm{lg}(R,h)}{\partial h}
- \frac{\Delta \gamma}{T}
\frac{\partial A_\mathrm{sg}(R,h)}{\partial h}
\nonumber \\  & = &
\left[ \frac{p_\mathrm{s}}{T} - \frac{p}{T} \right]
\pi \left[ 2 R h - h^2\right]
\nonumber \\ & & \mbox{ }
- \frac{\gamma_\mathrm{lg}  }{T}
2 \pi R
- \frac{\Delta \gamma}{T}
2 \pi [ R - h ] .
\end{eqnarray}

If the total entropy is a maximum simultaneously with respect to
curvature radius $R$ and apex height $h$,
then the left hand sides of Eqs~(\ref{Eq:ds/dR}) and (\ref{Eq:ds/dh})
are zero.
These two equations have simultaneous solution
\begin{equation}
\Delta \gamma =  -\gamma \frac{R-h}{R} = \gamma \cos \theta ,
\end{equation}
which is the contact angle condition,
and
\begin{equation}
p_\mathrm{s} = p + \frac{2 \gamma}{R} ,
\end{equation}
which is the Laplace-Young equation.
These two results hold whether or not the entropy is maximal
with respect to number
(ie.\ whether or not Eq.~(\ref{Eq:ds/dN}) is zero).

\begin{figure}[t!]
\centerline{
\resizebox{8.5cm}{!}{ \includegraphics*{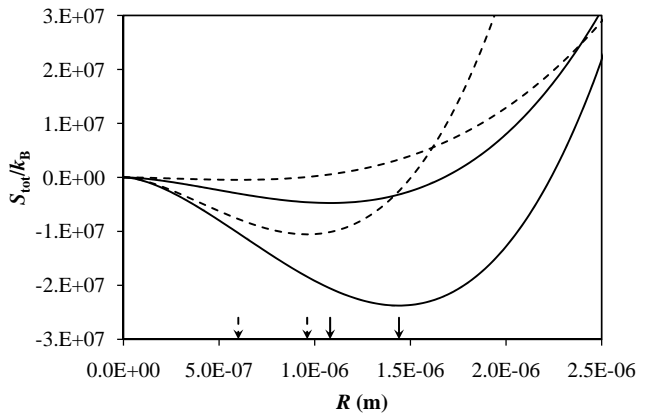} } }
\caption{\label{Fig:h2}
Total entropy of a mobile hemispherical air bubble in water
on a hydrophobic surface, $\Delta \gamma = -0.036\,$N/m,
following the mechanical equilibrium  path.
The solid curves are for a supersaturation ratio of $s=2$
and the dashed curves are for $s=2.5$.
The lower curve of each pair
uses the coexistence surface tension $\gamma^\dag = 0.072\,$N/m,
and the upper curve  of each pair uses the supersaturated surface tension
$\gamma(2) = 0.054\,$N/m, 
and $\gamma(2.5)=0.045\,$N/m. 
}
\end{figure}

\begin{table}[t]
\caption{\label{Tab:Hemi}
Geometry of a Mobile Hemispherical Bubble at the Critical Radius.
The difference in solid energies is $\Delta \gamma = -0.036\,$N/m.}
\begin{center}
\begin{tabular}{c c c c c c}
\hline\noalign{\smallskip}
$s$ &  $\gamma$ & $R_\mathrm{crit}$ & $r_\mathrm{crit}$ & $h_\mathrm{crit}$ &
$\theta_\mathrm{c}$ \\
    &     N/m   &   ($\mu$m)        &       ($\mu$m)    &      ($\mu$m)
& (deg.)\\
\hline \\
2.0 & 0.072$^\dag$ & 1.44  & 1.25  & 0.720 & 120 \\
2.0 & 0.054$^*$    & 1.08  & 0.805 & 0.360 & 132 \\
2.0 & 0.048$^\#$   & 0.960  & 0.635 & 0.240 & 139 \\ \\
2.5 & 0.072$^\dag$ & 0.960 & 0.831  & 0.480 & 120 \\
2.5 & 0.045$^*$    & 0.600 & 0.360 & 0.120 & 143 \\
2.5 & 0.036$^\#$   & 0.480 & 0.     & 0.   & 180 \\ \\
3.0 & 0.072$^\dag$ & 0.720 & 0.624  & 0.360 & 120 \\
3.0 & 0.036$^*$    & 0.360  & 0.    & 0.    & 180 \\
3.0 & 0.024$^\#$   & 0.240  & 0. & 0. & $>180$ \\
\hline
\end{tabular} \\
\flushleft
$^\dag$Coexistence value.\\
$^*$Supersaturated value using $s^\ddag = 5$.\\
$^\#$Supersaturated value using $s^\ddag= 4$.
\end{center}
\end{table}

In Fig.~\ref{Fig:h2}
the total entropy is plotted along the path of mechanical equilibrium,
which is equivalent to imposing the Laplace-Young equation
and the contact angle condition on the bubble as it grows.
It can be seen  that
the total entropy is a minimum at the critical radius.
The geometric parameters of a bubble at the critical radius
for these and other thermodynamic parameters are given in Table~\ref{Tab:Hemi}.
An optimum contact angle of 180$^\circ$ or greater
means that it is favorable for a planar gas layer
to form between the solid surface and the liquid phase
(drying transition).
The results in  Fig.~\ref{Fig:h2} confirm that a
hemispherical bubble mobile on a surface
has the same stability characteristics as a free spherical bubble,
which is to say that it remains unstable.

\subsection{Pinned Hemispherical Bubble}

For the case of a hemispherical bubble on a planar substrate,
if the contact rim of the bubble is mobile,
(ie.\ as it is when it satisfies the contact angle condition)
then its volume  increases with increasing radius of curvature,
just as for a free spherical bubble.
In both cases the critical radius is unstable.
However if the contact rim is pinned
then an increase in the radius of curvature decreases the volume,
and \emph{vice versa}.
In this case the critical radius becomes a point of stable equilibrium.

The physical origin of the stability is readily understood.
Following closely the discussion of the instability of a free bubble
given in the third last paragraph of the introduction,
at the critical radius of curvature for a pinned bubble,
a fluctuation to a larger radius of curvature
along the path of mechanical equilibrium
reduces the internal pressure and hence the gas chemical potential
inside the bubble.
There is now a thermodynamic gradient
that forces dissolved gas from the reservoir into the bubble,
which increases the internal pressure.
This creates a mechanical force imbalance,
which acts to increase the volume of the bubble
at the expense of the volume of the reservoir.
So far this is the same as for a free bubble and for a mobile adhering bubble.
But the difference for a pinned bubble is that
an increase in volume reduces the radius of curvature,
which counteracts the original fluctuation and restores
the bubble to its original state.
The converse occurs if the fluctuation is to a smaller radius of curvature.
Hence the critical radius is a point of stable equilibrium
for a pinned bubble.

This physical argument
that the pinned critical bubble is at an entropy maximum
can be readily confirmed by numerical calculation.
First one needs to derive the entropy derivatives,
the vanishing of which give the critical radius.
For a pinned (immobile) bubble attached to a hydrophobic surface,
the rim radius $r$ is constant.
Hence needs to use $r$ instead of $h$ as an independent variable
so that $r$ can be held fixed during the derivatives
with respect to $R$ and $N$.
Using the fact that $h = R- \sqrt{R^2-r^2}$,
the volume of the bubble may be written
\begin{eqnarray}
V & = &
\pi \left[ R h^2 - \frac{1}{3} h^3  \right]
 \\ \nonumber & = &
\pi \left[ \frac{2}{3} R^3  - \frac{2}{3}  R^2\sqrt{R^2-r^2}
- \frac{1}{3} r^2 \sqrt{R^2-r^2}
\right] ,
\end{eqnarray}
and the area of the liquid-vapor interface is
\begin{equation}
A_\mathrm{lg} =
2 \pi R \left[ R  - \sqrt{ R^2 - r^2}  \right] .
\end{equation}
The area of the solid-gas interface is of course
$A_\mathrm{sg} = \pi r^2  $.

The derivatives are
\begin{equation}
\frac{\partial V_\mathrm{s}(R,r)}{\partial R}
=
2 \pi  R^2  - \frac{4\pi R}{3} \sqrt{R^2-r^2}
- \frac{\pi [ 2 R^2 + r^2] R}{3\sqrt{R^2-r^2}} ,
\end{equation}
\begin{equation}
\frac{\partial A_\mathrm{lg}(R,r)}{\partial R}
=
4 \pi  R
-2 \pi \sqrt{R^2-r^2}
- \frac{2\pi R^2}{\sqrt{R^2-r^2}},
\end{equation}
\begin{equation}
\frac{\partial A_\mathrm{sg}(R,r)}{\partial R}
=
0 .
\end{equation}

With these, the derivatives of the total entropy  at fixed $r$ are
\begin{equation} 
\frac{\partial S_\mathrm{total}(N,R|r,T)}{\partial N}
= \frac{-\mu_\mathrm{s}}{T} + \frac{\mu}{T} ,
\end{equation}
and
\begin{eqnarray} 
\lefteqn{
\frac{\partial S_\mathrm{total}(N,R|r,T)}{\partial R}
} \nonumber \\
& = &
\left[ \frac{p_\mathrm{s}}{T} - \frac{p}{T} \right]
\frac{\partial V(R,r)}{\partial R}
- \frac{\gamma_\mathrm{lg}  }{T}
\frac{\partial A_\mathrm{lg}(R,r)}{\partial R} .
\end{eqnarray}
Hence at the bubble stationary point
when the radius derivative vanishes
(mechanical equilibrium for a pinned contact rim)
one has
\begin{eqnarray}
\overline p_\mathrm{s}
& = &
p
+
\gamma_\mathrm{lg}
\frac{\partial A_\mathrm{lg}(R,r)}{\partial R}
\frac{\partial R}{\partial V(R,r)}
\nonumber \\ & = &
p_\mathrm{r}
+
\gamma_\mathrm{lg}
\frac{4   \sqrt{R^2-r^2}  - 4 R + 2  r^2/R
}{ 2   R \sqrt{R^2-r^2} - 2 R^2 +  r^2 } .
\end{eqnarray}
If $ R \approx r$, then
$\overline p_\mathrm{s} \sim p_\mathrm{r} + 2 \gamma_\mathrm{lg}/R$,
which is the usual Laplace-Young expression.
More generally,
the left hand side is $ p_\mathrm{s} = N k_\mathrm{B} T /V(R,r)$.
Inserting this into the above
and taking the volume over to the right hand side
gives an explicit equation for $N(R)$
along the path of mechanical equilibrium.
This may be inserted into the expression for the total entropy
and the latter plotted as a function of the radius of curvature
for $R > r$.
For the case of diffusive equilibrium,
$N = s \rho_\mathrm{g}^\dag V(R,r)$.
Either the saturated surface tension $\gamma^\dag$
or the supersaturated surface tension  $\gamma(s)$ may be used.

\begin{figure}[t!]
\centerline{
\resizebox{8.5cm}{!}{ \includegraphics*{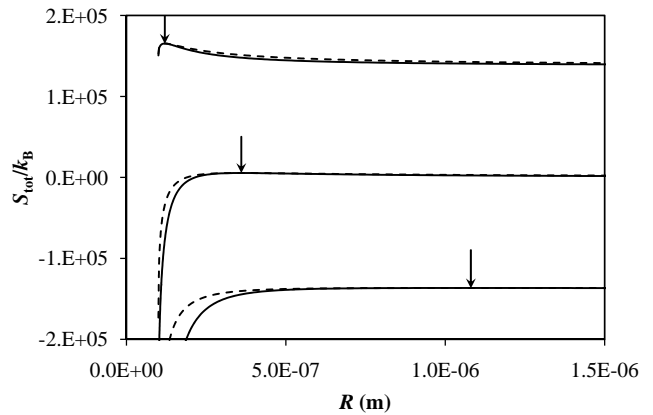} } }
\caption{\label{Fig:hp2}
Total entropy of a pinned hemispherical bubble
as a function of the radius of curvature
for  fixed contact rim radius $r=10^{-7}\,$m.
The solid curves follow the mechanical equilibrium  path
and dashed curves follow the diffusive equilibrium path.
The lower pair of curves are for
$s=2$ and $\gamma(s) = 0.054\,$N/m,
the middle pair use $s=3$ and $\gamma(s) = 0.036\,$N/m,
and the upper pair use $s=4$ and $\gamma(s) = 0.018\,$N/m
($s^\ddag = 5$ in all cases).
The arrows denote the  entropy maxima,
which are at the critical radii.
}
\end{figure}

\comment{
\begin{figure}[t!]
\centerline{
\resizebox{8.5cm}{!}{ \includegraphics*{Fig5old.eps} } }
\caption{\label{Fig:hp2}
Total entropy of a pinned hemispherical bubble
as a function of the radius of curvature for supersaturation $s=2$
and fixed contact rim radius $r=10^{-7}\,$m.
The solid curves follow the mechanical equilibrium  path,
and dashed curves follow the diffusive equilibrium path.
The lower pair of curves use the saturated surface tension,
$\gamma^\dag = 0.072\,$N/m,
and the upper pair of curves use the supersaturated surface tension,
$\gamma(s) = 0.054\,$N/m, ($s^\ddag = 5$).
The arrows denote the  entropy maxima,
which are at the critical radii.
}
\end{figure}
} 

Figure~\ref{Fig:hp2} shows the total entropy
as a function of the radius of curvature
when the contact rim is pinned at $r=10^{-7}\,$m.
Unlike the previous cases,
the total entropy is now a maximum at the critical radius.
The maximum is rather broad and extends
down almost to the pinned radius itself.
For supersaturations of $s=3$ and $s=4$, the total entropy maximum is positive,
which indicates that it is favorable to have a pinned bubble
compared to no bubble at all.

\begin{figure}[t!]
\centerline{
\resizebox{8.5cm}{!}{ \includegraphics*{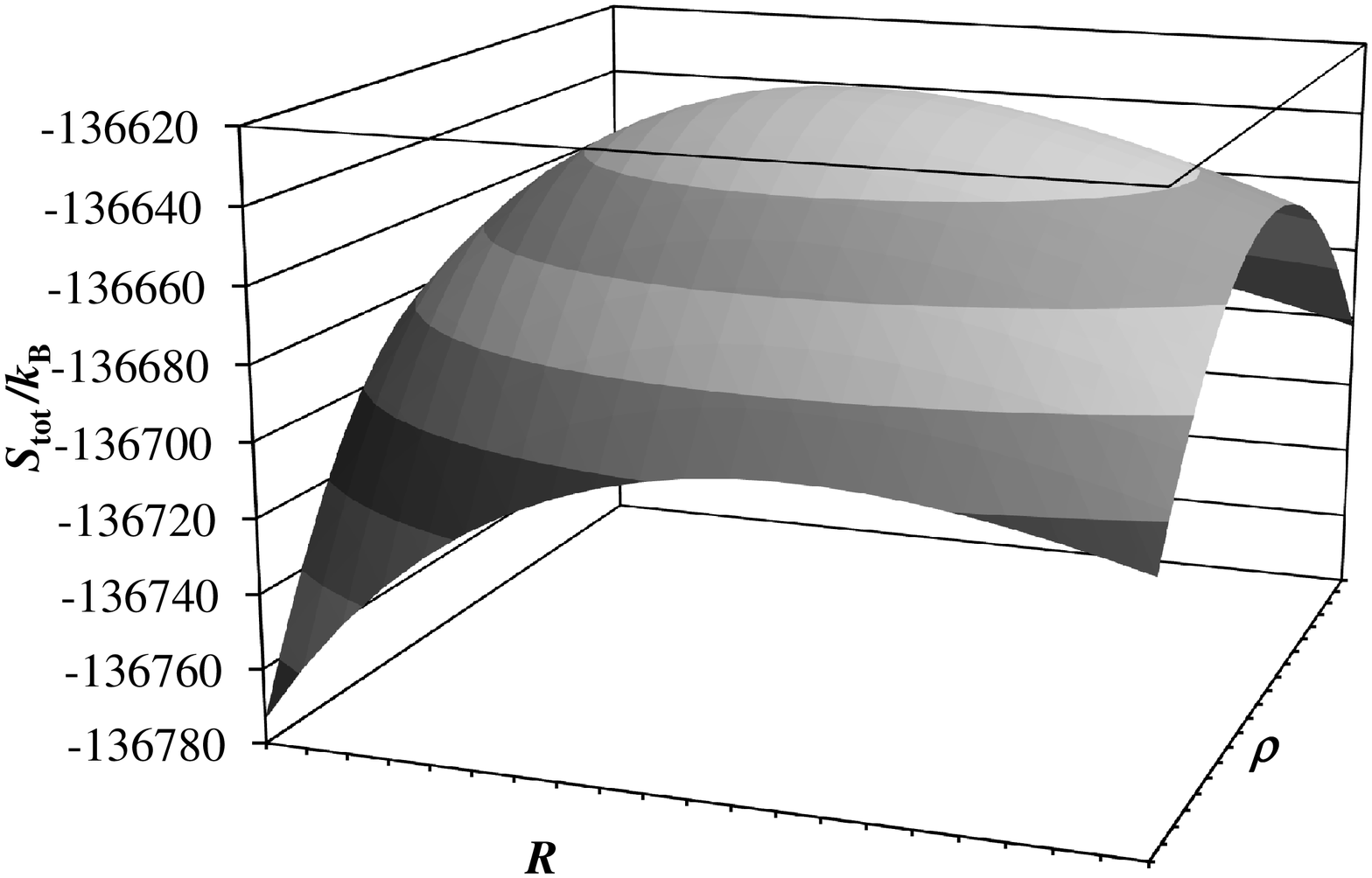} } }
\caption{\label{Fig:hp2-3D}
Total entropy of a pinned hemispherical bubble
($s=2$, $\gamma(s) = 0.054\,$N/m,  $r=10^{-7}\,$m)
in the neighborhood of the critical radius of curvature and density
($\pm 20$\% in each dimension).
}
\end{figure}

Figure~\ref{Fig:hp2-3D} is a surface plot of the total entropy
for the pinned hemispherical bubble
over the curvature radius-density plane.
It can clearly be seen that the critical radius and density
is a point of global entropy maximum.
In other words the pinned (immobile) bubble is stable.

%
\section{Conclusion}
\setcounter{equation}{0} 
%

In the introduction
four contradictions between classical nucleation theory
and nanobubbles were pointed out:
(1) in the classical theory bubbles are unstable
 whereas nanobubbles appear stable,
(2) the classical critical radius is 1--2 orders of magnitude larger
than the nanobubble radius of curvature,
(3) the classical internal pressure of nanobubbles
is  1--2 orders of magnitude larger than atmospheric pressure,
and (4) the macroscopic contact angle is significantly larger
than the measured nanobubble contact angle.

It was shown here that two non-classical ideas are necessary and sufficient
to account for the behavior of nanobubbles,
namely the dependence of surface tension on supersaturation,
and the pinning of the contact line of the nanobubble on the surface.

It has previously been pointed out
that bubbles can only be in diffusive equilibrium
with a supersaturated solution,\cite{Moody02}
and that the supersaturated surface tension
is smaller than the saturated surface tension.\cite{Moody03}
These facts reduce the size of the critical radius
and the magnitude of the internal pressure
from those predicted by classical nucleation theory.
This largely resolves the second and third problems enumerated above
since, depending upon the parameters,
it is possible to get quantitative agreement with the measured
radius of curvature of nanobubbles
and the deduced value of the surface tension\cite{Attard14} (see below)
for apparently reasonable values of the parameters
(actual supersaturation ratio of the solution
and the spinodal supersaturation ratio).
The origin of the supersaturation
has been variously attributed to the heating of the solution and the surface
by the laser diode, the piezo-electric crystal, and  exothermic
mixing during the exchange of ethanol for water,
and also to the entrainment of air by the surfaces.\cite{Attard14}
In any case, whatever the origin of the air,
once stable nanobubbles are present,
the inescapable thermodynamic conclusion is
that the solution is supersaturated.

The new contributions here on this point
were the explicit calculations for free and adsorbed bubbles,
and the mathematical proof
that the state-dependent surface tension on its own
can never give a stable nanobubble,
Appendix~\ref{Sec:Local-Super-Sat-Unstable}.

On the first point concerning stability,
it was confirmed that neither a free spherical bubble
nor a mobile hemispherical bubble on a  solid surface were stable.
However, if the bubble was pinned
with fixed contact rim, then it became stable.
The physical reason for the change in stability is that
whereas both a free spherical bubble and a mobile hemispherical bubble
have a volume that increases with increasing radius,
the volume of a pinned  hemispherical bubble
\emph{decreases} with increasing radius of curvature.
This is sufficient to confer stability.

The fourth point is  resolved by either or both of these non-classical notions.
The decrease in surface tension from the saturated value
of a macroscopic droplet will increase the contact angle
provided that the solid surface energies remain unchanged.
This observation was the origin of the estimate
of the surface tension of nanobubbles given in Ref.~\onlinecite{Attard14},
and it appears to be consistent with their measured radii of curvature.
However, if the nanobubble is pinned, then the contact angle will also
increase as the radius of curvature grows toward its critical value.

For example, for the case of a macroscopic contact angle of 120$^\circ$,
($\Delta \gamma = -0.036\,$N/m),
a mobile bubble using the saturated surface tension $\gamma^\dag = 0.072\,$N/m
has the same contact angle at the critical radius.
At $s=2$ and $ \gamma(s) = 0.054\,$N/m,
a mobile bubble has contact angle 132$^\circ$ at the critical radius.
A  bubble pinned at $r = 100\,$nm
has contact angle 176$^\circ$ for $\gamma^\dag = 0.072\,$N/m,
and 172$^\circ$ for $ \gamma(s) = 0.054\,$N/m,
each at the respective critical radius (entropy maximum).

Pinning evidently has a more dramatic effect on the contact angle
than supersaturation.
The fact that the measured contact angles for nanobubbles
are larger than that of macroscopic water droplets on the same surface
supports both non-classical notions
and on its own cannot decide between them.
Conversely,
the calculations in Ref.~\onlinecite{Attard14}
of the surface tension of nanobubbles from the measured contact angles
are probably not quantitatively reliable
because they neglect the possibility of pinning.

Two questions now arise:
is pinning the only mechanism that will confer stability to a bubble?
And what is the physical origin of pinning?

It has previously been shown
that a finite one-component  system leads to stable bubbles,\cite{Moody02}
and it is of interest whether similar results hold for nanobubbles
derived from air dissolved in water.
A model of a finite reservoir based upon the limited
diffusion of air in water in investigated in Appendix~\ref{Sec:Finite}.
This model does yield an entropy maximum and stable bubbles
at radii beyond the classical critical radius,
but only on unrealistically short time scales.
It is concluded that on experimental time scales
the reservoir of dissolved gas is effectively infinite due to diffusion.
Hence it appears that pinning is the only viable source of stability
for nanobubbles.

The most obvious sources of pinning are permanent
heterogeneities on the surface.
However, most measurements on nanobubbles are performed on surfaces
deliberately chosen to be smooth and chemically homogeneous.
This does not rule out random heterogeneities in any one case,
particularly when isolated nanobubbles are observed.
Because nanobubbles are so small,
a very low surface density of heterogeneities
may be sufficient to pin the contact line at a few points,
which would immobilize the whole rim.
Conversely, pinning at a few nanoscopic heterogeneities
likely has negligible effect on macroscopic bubbles or droplets.

Another possible source of pinning is the electric double layer repulsion
between neighboring nanobubbles.
The images of nanobubbles obtained by Tyrrell and Attard
\cite{Tyrrell01,Tyrrell02} show them to be close packed on the surface,
exactly as one would expect for the surface nucleation
from a supersaturated solution.
Further, the contact lines are often irregular and non-circular,
whereas a mobile bubble would be expected to have a circular contact line.
The nanobubbles are negatively charged  and their morphology
changes with pH.
Although the electric double layer repulsion
between neighboring nanobubbles would prevent
their contact lines from growing,
it would not prevent them from shrinking.
One concludes that it is probably not a general a source of stability.

\begin{figure}[t!]
\centerline{
\resizebox{8.5cm}{!}{ \includegraphics*{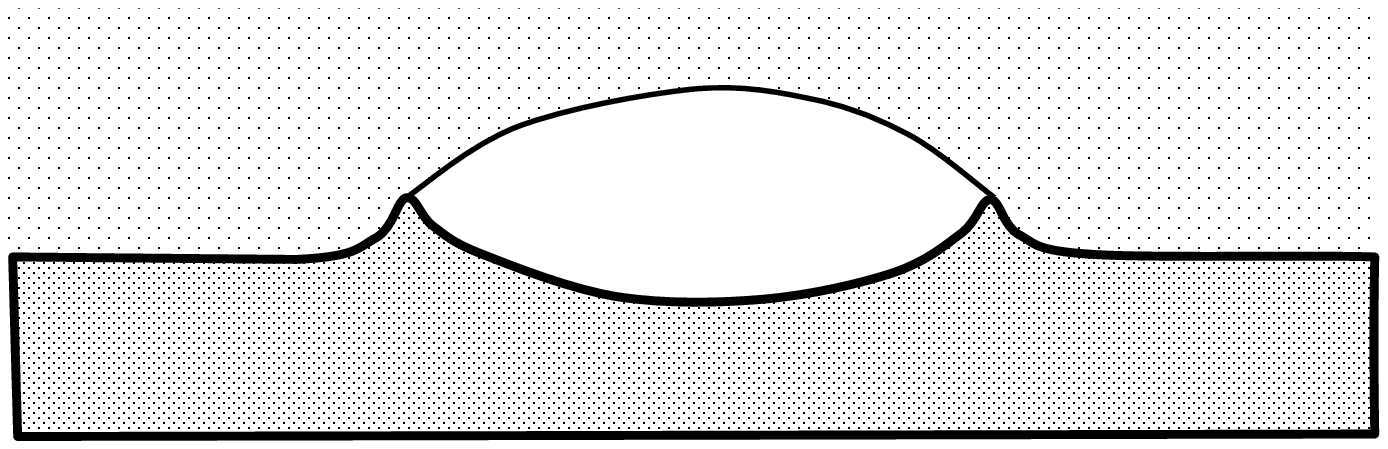} } }
\caption{\label{Fig:pin}
Sketch of a nanobubble pinned by its deformation of the solid surface
(not to scale).
}
\end{figure}

Even in the absence of permanent heterogeneities
or neighbor interactions,
it is possible that self-induced
pinning of the contact line may occur due to the elastic
or viscoelastic deformation of the substrate by the nanobubble,
as is sketched in Fig.~\ref{Fig:pin}.
The deformation is due to the twin effects of the excess pressure
inside the nanobubble
and the tension of the air-water interface applied at the contact line.
Depending upon the relaxation time of the substrate,
the barriers to relaxation,
and the driving forces for horizontal motion,
the contact rim may be effectively pinned at the deformation.
It is clear  that the cost of deformation
of the base and rim of the substrate favor a small contact radius,
whereas the cost of curvature of the contact rim favors a large contact radius.
The latter effect is much more significant for a nanobubble
than for a macroscopic bubble.


In conclusion,
this paper has shown that two non-classical notions
are necessary and sufficient to account for the size and stability
of nanobubbles:
the surface tension depends upon the level of supersaturation of the solution,
and pinning of the contact rim confers thermodynamic stability.

These appear to solve
the theoretical thermodynamic problems posed by nanobubbles.
They indicate the direction
for more experimental work to further characterize
nanobubbles and to establish that control over the phenomena
that is a prerequisite for industrial or technological exploitation.
Quantitative  measurements with controlled supersaturation,
either by means of de-aeration, over-gassing, or temperature control,
represent one line of enquiry.
Measurements with substrates of different deformability,
roughness, or polymer extensivity are also suggested
by the present results.



\appendix

\renewcommand{\theequation}{\Alph{section}.\arabic{equation}}

%
\section{Two-Component Formulation} \label{Sec:2cpt}
%

At the more fundamental level,
a  bubble and surrounding liquid
should be treated as two coexisting two-component mixtures,
one rich in air, and the other rich in water.
Let air be denoted by the subscript 1
and water by the subscript 2.
The total entropy is
\begin{eqnarray}
\lefteqn{
S_\mathrm{tot}(N_1,N_2,V|\mu_1,\mu_2,p,T)
} \nonumber \\
& = &
S_\mathrm{s}(N_1,N_2,T)
+ \frac{\mu_1}{T} + \frac{\mu_2}{T} - \frac{pV }{T}
- \frac{\gamma A}{T} .
\end{eqnarray}
Equilibrium is characterized by the vanishing of the derivatives,
namely
\begin{equation}
\overline \mu_\mathrm{s1} = \mu_1 , \;\;
\overline \mu_\mathrm{s2} = \mu_2
\end{equation}
and
\begin{equation}
\overline p_\mathrm{s} = p + \frac{2 \gamma }{R} .
\end{equation}

At saturation one has
\begin{equation}
p = p_1^\dag + p_2^\dag ,
\end{equation}
where $p$ is atmospheric pressure.
For a supersaturation ratio of $s$,
the liquid solution phase (ie.\ the reservoir)
is in equilibrium 
with a gas phase of partial air pressure $s p_1^\dag$
and partial water vapor pressure of $p_2^\dag$.
When the bubble is in  diffusive equilibrium with the solution,
the sum of these must give the internal pressure of the bubble.
Hence the critical radius is given by
\begin{eqnarray}
R_\mathrm{crit} & = &
\frac{2 \gamma}{ s  p_1^\dag + p_2^\dag - p }
 \nonumber \\ & = &
\frac{2 \gamma}{ (s -1)p_1^\dag}.
\end{eqnarray}
The difference between this and the one-component formulation
is that the partial vapor pressure of air at saturation,
$p_1^\dag \approx 0.8 p$,
has been replaced by atmospheric pressure, $p$.

%
\section{Unstable Critical Radius} \label{Sec:Unstab-Rc}
%

As was  shown explicitly in Fig.~\ref{Fig:b2} above,
the critical radius is a saddle point,
which is an unstable equilibrium point.
This may be proven mathematically
by showing that a bubble does not satisfy
the general stability requirements.

\subsection{General Stability Condition}

Since the probability of a fluctuation in number or volume
is proportional to the exponential of the entropy,
two conditions necessary  for stability are that the
corresponding second derivatives of the entropy are negative,
\begin{equation}
S_{NN} < 0 , \mbox{ and } S_{VV}< 0,
\end{equation}
at least at a stable stationary point.
More generally,
the necessary and sufficient condition for stability
is that both eigenvalues of the Hessian matrix of second derivatives
are negative.
The eigenvalues are determined from the characteristic equation,
\begin{eqnarray}
\lefteqn{
\left| \begin{array}{cc}
S_{NN} - \lambda & S_{NV} \\
S_{VN} & S_{VV} - \lambda
\end{array} \right|
} \nonumber \\
& = &
[S_{NN} - \lambda] [S_{VV} - \lambda] - S_{VN}^2
 \nonumber \\ & = &
 \lambda^2 - \lambda \left[S_{NN}+S_{VV} \right] +
\left[ S_{NN} S_{VV} - S_{NV}^2 \right] ,
\end{eqnarray}
whose roots are
\begin{eqnarray}
\lambda_\pm & = &
\frac{1}{2} \left\{ \rule{0.cm}{.5cm}
\left[S_{NN}+S_{VV} \right]
\right.  \\  && \left. \mbox{ }
\pm \sqrt{ \left[S_{NN}+S_{VV} \right]^2
- 4 \left[ S_{NN} S_{VV} - S_{NV}^2 \right] }
\right\} .
\nonumber
\end{eqnarray}
The argument of the square root is always non-negative.
The only way for both eigenvalues to be negative
is if the square root is smaller than the magnitude of the first term,
which occurs when the second bracketed term in the square root is positive.
Hence in addition to the negativity of the two pure second derivatives,
stability is ensured by the positivity of the determinant of the
Hessian matrix,
\begin{equation}
S_{NN} S_{VV} - S_{NV}^2 >  0.
\end{equation}

\subsection{Bubble Instability}

Specifically for a bubble,
the second derivatives of the total entropy are
\begin{equation}
S_{NN} = \frac{-k_\mathrm{B}}{N} ,
\end{equation}
\begin{eqnarray}
S_{VV} & = &
\frac{-\rho k_\mathrm{B}}{V}
+ \frac{2 \gamma}{T R^2} \frac{1}{4 \pi R^2}
\nonumber \\ & = &
\frac{-p_\mathrm{s}  }{VT}
+ \frac{2 \gamma}{3 V T R}  ,
\end{eqnarray}
and
\begin{equation}
S_{NV} =
\frac{ k_\mathrm{B}}{V} .
\end{equation}
Clearly $S_{NN}$ is always negative.
At the equilibrium point,
the Laplace-Young equation shows that
$\overline S_{VV}
=
- ({p_\mathrm{r} }/{VT})
- ({2 \gamma }/{VT\overline R})
+ ({2 \gamma}/{3 V T \overline R}) $,
which is negative.
These two conditions are necessary but not sufficient conditions
for the equilibrium point to be stable.

The determinant of the Hessian matrix is
\begin{eqnarray}
S_{NN} S_{VV} - S_{NV}^2
 & = &
\frac{ k_\mathrm{B}^2 }{V^2 }
\left\{
\frac{-1}{\rho}
\left[ -\rho
+ \frac{2 \gamma}{3 k_\mathrm{B} T R} \right]
-1
\right\}
\nonumber \\ & = &
\frac{ - k_\mathrm{B}^2 }{V^2 }
\frac{2 \gamma}{3 \rho k_\mathrm{B} T R} ,
\end{eqnarray}
which is always negative.
As shown above,
it is necessary for stability that this determinant be positive.

One concludes from this that the bubble's stationary point,
the critical radius $ R_\mathrm{crit}$ and density $ \rho_\mathrm{crit}$,
is a saddle point,
which is a point of unstable equilibrium.

%
\section{Local Supersaturation} \label{Sec:Local-Super-Sat}
%

\subsection{Critical Radius for Local Supersaturation}

The simulations that justify the formula
for the  supersaturated surface tension
were carried out with diffusive
equilibrium established across the interface.
In homogeneous nucleation theory it is more traditional
to plot the total entropy (equivalently free energy)
as a function of radius along the path of mechanical equilibrium.
In this case diffusive equilibrium does not hold except
at the critical radius.
One could argue that there is always a local diffusive equilibrium
in the immediate locality of the bubble air-water interface,
and that this is what determines the value of the surface tension.
In this case one would use $\gamma(s_\mathrm{local})
\equiv \gamma(\rho)$,
where
$ s_\mathrm{local}(\rho)
\equiv \rho/\rho_\mathrm{g}^\dag
\equiv p_\mathrm{s}/p_\mathrm{g}^\dag$,
for the  bubble  on the path of mechanical equilibrium.

If one uses this bubble-dependent model for the supersaturated surface tension,
the total entropy is
\begin{eqnarray}
S_\mathrm{total}(N,R,T)
& = &
S_\mathrm{s}(N,V,T)
+ \frac{\mu_\mathrm{r} N}{T}
- \frac{p_\mathrm{r}V}{T}
\nonumber \\ & & \mbox{ }
- \frac{\gamma(\rho) A }{T} ,
\end{eqnarray}
with $\rho = N/V$
and $ p_\mathrm{r} = \rho_\mathrm{g}^\dag k_\mathrm{B}T$.
For full generality, and to make an important point,
it will \emph{not} here be assumed that
$\gamma(\rho)$ is a linear function of density.

The number derivative is
\begin{equation} \label{Eq:ds/dN3}
\frac{\partial S_\mathrm{total}}{\partial N}
 =
\frac{-\mu_\mathrm{s}}{T} + \frac{\mu_\mathrm{r}}{T}
- \gamma'(\rho) \frac{A }{V T} .
\end{equation}
When this derivative vanishes,
the chemical potential inside the bubble is higher than that outside.

The derivative with respect to radius is
\begin{eqnarray}\label{Eq:ds/dR2}
\frac{\partial S_\mathrm{total}}{\partial R}
& = &
\left[ \frac{p_\mathrm{s}}{T} - \frac{p_\mathrm{r}}{T} \right]
\frac{\partial V(R)}{\partial R}
\nonumber \\  & & \mbox{ }
- \frac{\gamma  }{T}
\frac{\partial A(R)}{\partial R}
- \frac{\gamma'(\rho) A }{T}
\frac{\partial \rho(R,N)}{\partial R}
\nonumber \\  & = &
\left[ \frac{p_\mathrm{s}}{T} - \frac{p_\mathrm{r}}{T} \right]
4 \pi R^2
\nonumber \\ & & \mbox{ }
- \frac{ \gamma(\rho)}{T}   8 \pi R
+ \gamma'(\rho)
\frac{ 12 \pi R \rho }{T } .
\end{eqnarray}
This vanishes when
\begin{equation}
\overline p_\mathrm{s}
=
 p_\mathrm{r}
+  \frac{ 2 \gamma(\rho) }{R}
- \frac{ 3 \rho \gamma'(\rho) }{R }  .
\end{equation}
Like the usual Laplace-Young equation,
this says that the internal pressure of the air bubble
is larger than the external pressure of the liquid,
since $\gamma(\rho) \ge 0$,
and $\gamma'(\rho) < 0$.
Since the usual Laplace-Young equation uses the coexistence
value for the surface tension,
$\gamma^\dag \equiv \gamma(\rho^\dag)
\le \gamma(\rho)$,
whether the internal pressure predicted by the present
theory is larger or smaller than that predicted
by the constant surface tension theory
depends on the specific surface tension function.

\subsection{Instability of the Critical Radius for Local Supersaturation}
\label{Sec:Local-Super-Sat-Unstable}

At the bubble critical radius
$( \rho_\mathrm{crit} ,  R_\mathrm{crit})$,
both derivatives vanish, $S_N( \rho_\mathrm{crit} ,  R_\mathrm{crit}) = 0$
and $S_V( \rho_\mathrm{crit} ,  R_\mathrm{crit}) = 0$.
Individually they give  a non-linear function for the optimum density
\begin{equation}
\overline \rho =
s \rho^\dag e^{ - 3 \gamma'(\overline \rho)
/k_\mathrm{B}T  R},
\end{equation}
and
\begin{equation}
\overline \rho  =
\rho^\dag
+ \frac{2 \gamma(\overline \rho)
- 3 \overline \rho \gamma'(\overline \rho)
}{k_\mathrm{B} T R}  .
\end{equation}
The simultaneous solution of these gives the critical radius and density.
If the derivative of the surface tension is negative,
$ \gamma'(\overline \rho)  < 0$,
then the first of these says that the gas density inside the bubble
is greater than the supersaturation vapor density.
The second says that the  gas density inside the bubble
is greater than the coexistence density,
both terms of the correction being positive.

Solving instead for the critical radius one obtains
\begin{equation}
\overline R =
\frac{ - 3 \gamma'(\overline \rho) }{ k_\mathrm{B}T }
\ln\frac{\overline \rho}{s \rho^\dag } ,
\end{equation}
and
\begin{equation}
\overline R =
\frac{2 \gamma(\overline \rho)
- 3 \overline \rho \gamma'(\overline \rho)
}{k_\mathrm{B} T[ \overline \rho - \rho^\dag] }  .
\end{equation}
The second of these gives
$R(\rho)$  for $\rho \in [\rho^\dag_\mathrm{g},\rho^\ddag_\mathrm{g}]$
along the path of mechanical equilibrium.

Since the critical radius must vanish at the spinodal,
$\overline R(\rho^\ddag_\mathrm{g}) = 0$,
this says that
\begin{equation}
\gamma(\rho^\ddag_\mathrm{g})
= \frac{3}{2} \rho^\ddag_\mathrm{g} \gamma'(\rho^\ddag_\mathrm{g}) .
\end{equation}
It is not possible to satisfy this in the linear model,
since the left hand side is positive or zero
and the right hand side is strictly negative.
Hence the local supersaturation-dependent surface tension
must be a non-linear function of the internal gas density.
Since in general
$\gamma'(\rho^\dag_\mathrm{g}) < 0$,
and  this result says that
$\gamma'(\rho^\ddag_\mathrm{g}) \ge 0$,
one must have that
\begin{equation} \label{Eq:g''>0}
\gamma''(\rho)  \ge 0 ,
\end{equation}
assuming that the interval $[\rho^\dag_\mathrm{g},\rho^\ddag_\mathrm{g}]$
is relatively small.
This result holds only for the local supersaturation-dependent surface tension.
It will now be shown that this result
precludes the critical radius
from being a point of stable equilibrium.

The second derivatives of the total entropy are
\begin{equation}
S_{NN} = \frac{-k_\mathrm{B}}{N}
- \frac{3 \gamma''(\rho)  }{V  TR},
\end{equation}
\begin{eqnarray}
S_{VV} & = &
\frac{-\rho k_\mathrm{B}}{V}
+ \frac{2 \gamma(\rho) - 3 \rho \gamma'(\rho) }{T R^2}
 \frac{1}{4 \pi R^2}
\nonumber \\ & & \mbox{ }
- \frac{  \gamma'(\rho) + 3  \rho \gamma''(\rho)}{TR}
\frac{ \rho  }{V}
\nonumber \\ & = &
\frac{-\rho k_\mathrm{B}}{V}
+ \frac{2 \gamma(\rho)
- 6 \rho \gamma'(\rho)
-9 \rho^2 \gamma''(\rho)
}{3 V T R }
 ,
\end{eqnarray}
and
\begin{equation}
S_{NV} =
\frac{ k_\mathrm{B}}{V}
+ \frac{3 \gamma'(\rho)  }{T R^2} \frac{1}{4\pi R^2}
+ \frac{3 \gamma''(\rho)  }{TR} \frac{\rho}{V} .
\end{equation}

Clearly $\overline S_{NN}$ is  negative if
\begin{equation}
\gamma''(\overline \rho) >
\frac{ -k_\mathrm{B}T \overline R}{3\overline \rho} .
\end{equation}
Although this bound \emph{could} be satisfied by
$\gamma''(\overline \rho)=0$,
this possibility is ruled out by another bound below.

Also $\overline S_{VV}$ is  negative if
\begin{equation}
0 >
-\overline \rho k_\mathrm{B}T
+ \frac{2 \gamma(\overline \rho)
- 6 \overline \rho \gamma'(\overline \rho)
-9 \overline\rho^2 \gamma''(\overline\rho)
}{3  \overline R } ,
\end{equation}
which can be written as
\begin{equation}
\gamma''(\overline\rho)
>
\frac{
-  3 \overline R \overline \rho k_\mathrm{B}T
+ 2 \gamma(\overline \rho)
- 6 \overline \rho \gamma'(\overline \rho)
}{ 9 \overline \rho^2 }   ,
\end{equation}
or as
\begin{equation}
 \gamma'(\overline \rho)
>
\frac{
-3  \overline R  \overline \rho k_\mathrm{B}T
+ 2 \gamma(\overline \rho)
-9 \overline\rho^2 \gamma''(\overline\rho)
}{ 6 \overline \rho } .
\end{equation}

The determinant should be evaluated at the bubble critical point,
which is denoted by an over-line.
For typographical simplicity, the density argument of the surface tension
will be suppressed.
One has
\begin{eqnarray}
\frac{V^2}{k_\mathrm{B}^2} S_{NN} S_{VV}
& = &
\left\{ \frac{-1}{\rho}
- \frac{3 \gamma''  }{R k_\mathrm{B} T} \right\}
\nonumber \\ & & \mbox{ } \times
\left\{
-\rho
+ \frac{2 \gamma
- 6 \rho \gamma'
- 9 \rho^2 \gamma''
}{3 R k_\mathrm{B} T}
\right\}
\nonumber \\ & = &
1 - \frac{2 \gamma
- 6 \rho \gamma'
- 9 \rho^2 \gamma''
}{3 R \rho k_\mathrm{B} T}
+ \frac{3 \rho \gamma''  }{R k_\mathrm{B} T}
\nonumber \\ & & \mbox{ }
-
\frac{2 \gamma \gamma''
- 6 \rho \gamma' \gamma''
- 9 \rho^2 \gamma''^2
}{(R k_\mathrm{B} T)^2}
\end{eqnarray}
and
\begin{eqnarray}
\frac{V^2}{k_\mathrm{B}^2} S_{NV}^2
& = &
\left\{
1
+ \frac{\gamma'  }{ R k_\mathrm{B} T}
+ \frac{3 \rho \gamma''  }{ R k_\mathrm{B}T}
\right\}^2
 \\ \nonumber & = &
1 +
\frac{2 \gamma' + 6 \rho \gamma'' }{ R k_\mathrm{B} T}
+
\frac{ 6 \rho \gamma' \gamma''
+\gamma'^2 + 9 \rho^2 \gamma''^2
}{ (R k_\mathrm{B} T)^2} .
\end{eqnarray}
For a stable bubble critical point,
the determinant should be positive
\begin{eqnarray}
0 & < &
\frac{-2 \overline \gamma
+ 6 \overline\rho \overline\gamma'
}{3 \overline R \overline\rho k_\mathrm{B} T}
-
\frac{2 \overline\gamma \overline\gamma''
- 6 \overline\rho \overline\gamma' \overline\gamma''
-9 \overline\rho^2 \overline\gamma''^2
}{(\overline R k_\mathrm{B} T)^2}
\nonumber \\  & & \mbox{ }
-\frac{2 \overline \gamma'
+ 6 \overline\rho \overline\gamma'' }{ \overline R k_\mathrm{B} T}
-
\frac{ 6 \overline\rho \overline\gamma'
 \overline\gamma''
+ \overline\gamma'^2
+ 9 \overline\rho^2 \overline\gamma''^2
}{ (\overline R k_\mathrm{B} T)^2}
%
\nonumber \\ & = &
\frac{-2 \overline \gamma
- 18 \overline \rho^2 \overline \gamma''
 }{3 \overline R \overline \rho k_\mathrm{B} T}
-
\frac{2 \overline \gamma \overline \gamma''
+ \overline \gamma'^2
}{(\overline R k_\mathrm{B} T)^2} .
\end{eqnarray}
This gives
\begin{equation}
\overline \gamma'^2
<
(\overline R k_\mathrm{B} T)^2
\left[
\frac{-2 \overline \gamma
- 18 \overline \rho^2 \overline \gamma''
 }{3 \overline R \rho k_\mathrm{B} T}
-
\frac{2 \overline \gamma \overline\gamma''
}{(\overline R k_\mathrm{B} T)^2}
\right] .
\end{equation}
This has a solution if, and only if,
the bracketed term is positive, which means that
\begin{eqnarray}
\frac{-2 \overline \gamma }{3 \overline R \rho k_\mathrm{B} T}
& > &
\frac{2 \overline \gamma \overline\gamma''
}{(\overline R k_\mathrm{B} T)^2}
+
\frac{ 18 \overline \rho^2 \overline \gamma''
 }{3 \overline R \rho k_\mathrm{B} T}
\nonumber \\ & = &
\frac{ 2 \overline \gamma
+ 6 \overline \rho \overline R k_\mathrm{B} T
}{(\overline R k_\mathrm{B} T)^2}
\overline\gamma'' ,
\end{eqnarray}
or
\begin{eqnarray}
\overline \gamma'' & < &
\frac{-2 \overline \gamma }{3 \rho \overline R  k_\mathrm{B} T}
\frac{(\overline R k_\mathrm{B} T)^2
}{ 2 \overline \gamma
+ 6 \overline \rho \overline R k_\mathrm{B} T  }
\nonumber \\ & = &
\frac{-  \overline R k_\mathrm{B} T /3 \overline \rho
}{ 1 +
3 \overline \rho \overline R k_\mathrm{B} T /\overline \gamma } .
\end{eqnarray}
Combining this with the condition found above for $\overline S_{NN} < 0$,
$\overline \gamma'' >
{ -\overline R k_\mathrm{B}T }/{3\overline \rho}$,
gives
\begin{equation}
\frac{\overline R k_\mathrm{B} T /3 \overline \rho
}{ 1 +
3 \overline \rho \overline R k_\mathrm{B} T /\overline \gamma }
< - \overline \gamma'' <
\frac{ \overline R k_\mathrm{B}T }{3\overline \rho} .
\end{equation}
This rules out the possibility that $\overline \gamma'' = 0$,
which is to say that the local supersaturated
surface tension cannot be a linear function of density
if a bubble is to be stable.
This agrees with the conclusion already made above on different grounds.

This result shows that the second derivative of the surface tension
has to be negative if the free bubble is to be stable.
But Eq.~(\ref{Eq:g''>0}) showed that
the second density derivative of the surface tension
had to be positive $\gamma''(\rho) > 0$.
These two results prove
that it is not possible to confer thermodynamic stability on a bubble
at the critical radius by invoking
a local supersaturation  surface tension.


%
\section{Finite Reservoir} \label{Sec:Finite}
%

In order to model the finite diffusion of air in water,
one can consider two reservoirs:
an infinite volume reservoir of atmospheric pressure $p$,
and a reservoir of fixed volume $V_\mathrm{r}$.
The total volume $V_\mathrm{tot} = V + V_\mathrm{r}$
is variable and can exchange with the infinite volume pressure reservoir.
The total number of of air molecules is fixed,
$N_\mathrm{tot} = N + N_\mathrm{r} = s p V_\mathrm{r}/K$,
where $K=2.3\times 10^{-19}\,$J is Henry's constant for air in water.
The idea is that due to limited diffusion over a specified time scale,
only air dissolved in the water of the fixed volume reservoir
can  exchange with the bubble.

One can define the fixed reservoir volume in terms of a diffusion length
\begin{equation}
V_\mathrm{r} = \frac{4 \pi l_D^3 }{3} , \;\;
l_D \equiv \sqrt{D t} ,
\end{equation}
where $D = 2 \times 10^{-9} \,$m$^2$/s
is the diffusion constant for air and water,
and $t$ is some experimental time scale.

The spherical bubble or sub-system is characterized by
number $N$, volume $V=4\pi R^3/3$, and area $A=4\pi R^2$.
With these the total entropy is
\begin{eqnarray}
S_\mathrm{total}
& = &
S_\mathrm{s}(N,V,T)
+ S_\mathrm{r}(N_\mathrm{r},V_\mathrm{r},T)
- \frac{\gamma}{T} A
\nonumber \\ && \mbox{ }
- \frac{p}{T}V_\mathrm{tot}
+ \mbox{const}.
\end{eqnarray}
The constant is chosen to make the entropy vanish at zero radius.

As usual the sub-system (bubble) is an ideal gas
\begin{equation}
S_\mathrm{s}(N,V,T)
=
k_\mathrm{B} N \left[ 1 - \ln \frac{N \Lambda^3}{V} \right] .
\end{equation}
The fixed volume reservoir containing the dissolved air
can be treated as an ideal gas in a uniform  external field
\begin{equation}
S_\mathrm{r}(N_\mathrm{r},V_\mathrm{r},T)
=
k_\mathrm{B} N_\mathrm{r}
\left[ 1 - \ln \frac{N_\mathrm{r} \Lambda^3}{V_\mathrm{r}} \right]
- \frac{ N_\mathrm{r} \varepsilon }{T} .
\end{equation}
Here $\varepsilon$ is the molecular solvation energy.
The effective chemical potential that corresponds to this is
\begin{equation}
\mu_\mathrm{r}
= \varepsilon + k_\mathrm{B} T \ln \rho_\mathrm{sol}\Lambda^3 ,
\end{equation}
where $\rho_\mathrm{sol}  = N_\mathrm{r}/V_\mathrm{r}$
is the concentration of dissolved gas.
Henry's law may be written for the concentration
of dissolved gas in the form
\begin{equation}
\rho_\mathrm{sol}  = K^{-1} p
= K^{-1} k_\mathrm{B} T  \Lambda^{-3} e^{\mu/k_\mathrm{B} T} ,
\end{equation}
or
\begin{equation}
\mu =
k_\mathrm{B} T \ln \frac{K \rho_\mathrm{sol} \Lambda^{3}}{k_\mathrm{B} T} .
\end{equation}
Recall that the saturation air vapor pressure
has been taken to equal the external atmospheric pressure.
Equating these two expressions for the chemical potential yields
\begin{equation}
\varepsilon =
k_\mathrm{B} T \ln \frac{ K }{ k_\mathrm{B} T }.
\end{equation}

Because the reservoir volume $V_\mathrm{r}$
is fixed and independent of the bubble,
the derivative of the total entropy with respect to radius
vanishes when the Laplace-Young equation is satisfied,
\begin{equation}
\overline p_\mathrm{s} = p_\mathrm{ext} + \frac{2 \gamma}{R} .
\end{equation}
This gives the path of mechanical equilibrium.

\begin{figure}[t!]
\centerline{
\resizebox{8.5cm}{!}{ \includegraphics*{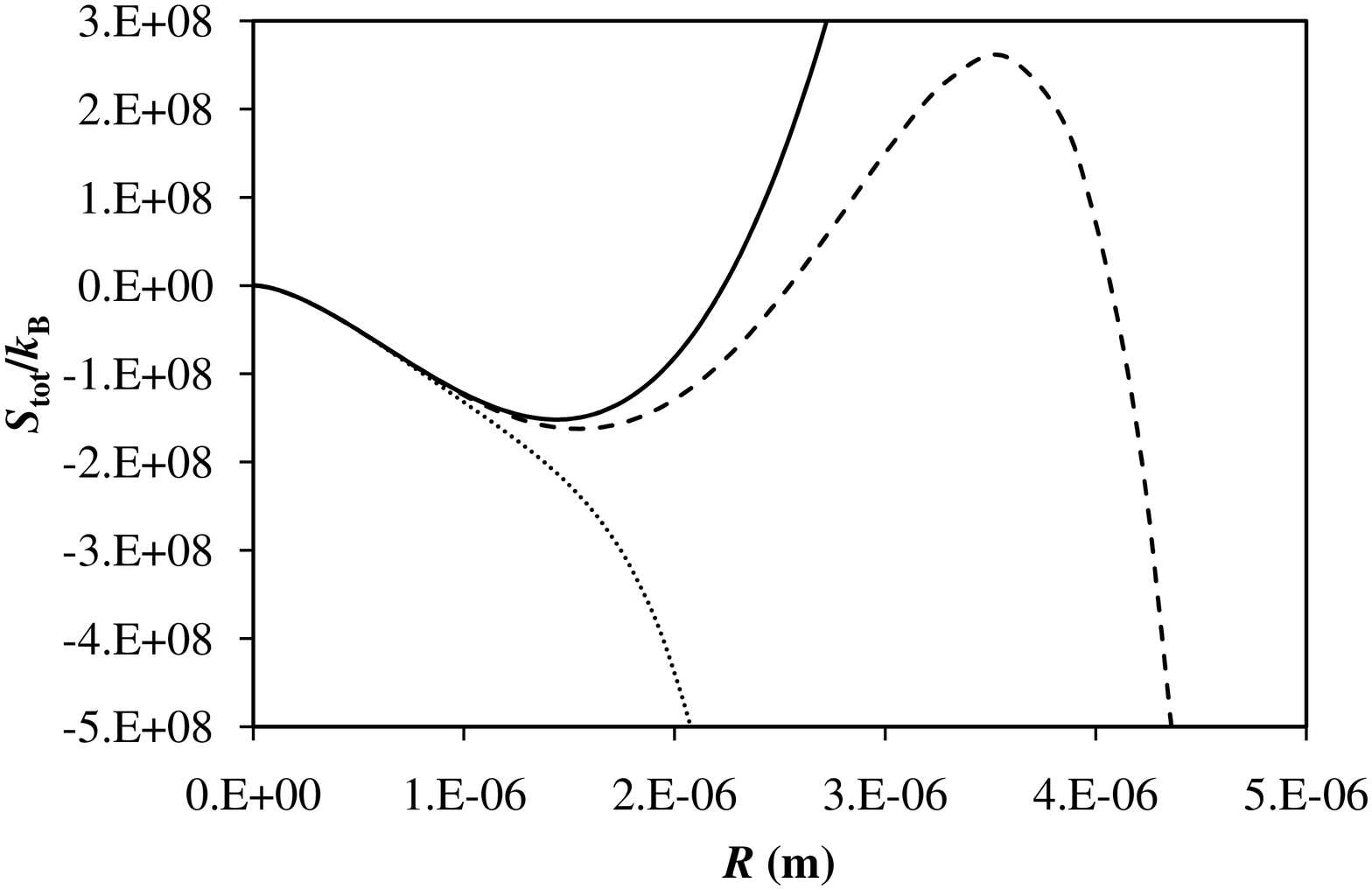} } }
\caption{\label{Fig:bf2}
Total entropy of a spherical bubble
in a finite reservoir on the path of mechanical equilibrium
($s=2$, $\gamma^\dag = 0.072\,$N/m).
The solid curve is for an infinite reservoir,
the dashed curve is for a diffusion time $t=0.035\,$s,
and the dotted curve is for  $t=0.01\,$s.
}
\end{figure}

Figure~\ref{Fig:bf2} plots the entropy
using this model for a finite, diffusion-linked reservoir.
The results are typical of the model.
For small bubbles the reservoir is effectively infinite
and all the curves coincide.
For larger bubbles and small reservoirs
(ie.\ short experimental time scales),
the total entropy monotonically decreases and there is no critical radius.
For larger reservoirs
(ie.\ longer experimental time scales),
there is a critical radius
and beyond this a stable radius where the entropy is a maximum.
The decrease in entropy beyond this maximum is due to the depletion
of the reservoir.
Obviously, the radius of maximum entropy will increase with
increasing reservoir size.
It should be noted that the typical lifetime observed for nanobubbles
is on the order of $10^3$--$10^4\,$s.

\newpage

\end{document}